# MULTILAYER PERCEPTRON GUIDED KEY GENERATION THROUGH MUTATION WITH RECURSIVE REPLACEMENT IN WIRELESS COMMUNICATION (MLPKG)

Arindam Sarkar[1] and J. K. Mandal[2]

[1]Department of Computer Science & Engineering, University of Kalyani, W.B, India
arindam.vb@gmail.com
[2]Department of Computer Science & Engineering, University of Kalyani, W.B, India
jkm.cse@gmail.com

## *ABSTRACT*

*In this paper, a multilayer perceptron guided key generation for encryption/decryption (MLPKG) has been proposed through recursive replacement using mutated character code generation for wireless communication of data/information. Multilayer perceptron transmitting systems at both ends accept an identical input vector, generate an output bit and the network are trained based on the output bit which is used to form a protected variable length secret-key. For each session, different hidden layer of multilayer neural network is selected randomly and weights or hidden units of this selected hidden layer help to form a secret session key. The plain text is encrypted using mutated character code table. Intermediate cipher text is yet again encrypted through recursive replacement technique to from next intermediate encrypted text which is again encrypted to form the final cipher text through chaining , cascaded xoring of multilayer perceptron generated session key. If size of the final block of intermediate cipher text is less than the size of the key then this block is kept unaltered. Receiver will use identical multilayer perceptron generated session key for performing deciphering process for getting the recursive replacement encrypted cipher text and then mutated character code table is used for decoding. Parametric tests have been done and results are compared in terms of Chi-Square test, response time in transmission with some existing classical techniques, which shows comparable results for the proposed technique.*

## 1. INTRODUCTION

In recent times wide ranges of techniques are developed to protect data and information from eavesdroppers [1, 2, 3, 4, 5, 6, 7, 8, 9]. These algorithms have their virtue and shortcomings. For Example in DES, AES algorithms [1] the cipher block length is nonflexible. In NSKTE [4], NWSKE [5], AGKNE [6], ANNRPMS [7] and ANNRBLC [8] technique uses two neural networks one for sender and another for receiver having one hidden layer for producing synchronized weight vector for key generation. Now attacker can get an idea about sender and receiver’s neural machines because for each session architecture of neural machine is static. In NNSKECC algorithm [9] any intermediate blocks throughout its cycle taken as the encrypted block and this number of iterations acts as secret key. Here if n number of iterations are needed for cycle formation and if intermediate block is chosen as an encrypted block after $n/2^{th}$ iteration then exactly same number of iterations i.e. n/2 are needed for decode the block which makes easier the attackers life. To solve these types of problems in this paper we have proposed a multilayer perceptron guided encryption technique in wireless communication.

The organization of this paper is as follows. Section 2 of the paper deals with the problem domain and methodology. Proposed Multilayer Perceptron based key generation has been discussed in section 3. Character code table generation technique is given in section 4. Recursive replacement encryption and example of encryption has been presented in section 5 and 6 respectively. Section 7 and 8 deals with recursive replacement decryption and example of decryption method. Complexity analysis of the technique is given in section 9. Experimental results are described in section 10. Analysis of the results presented in section 11. Analysis regarding various aspects of the technique has been presented in section 12. Conclusions and future scope are drawn in section 13 and that of references at end.

## 2. Problem Domain and Methodology

In security based communication the main problem is distribution of key between sender and receiver. Because at the time of exchange of key over public channel intruders can intercept the key by residing in between them. This particular problem has been addressed and a technique has been proposed technique addressed this problem. These are presented in section 2.1 and 2.2 respectively.

### 2.1. Man-In-The-Middle Attack

Intruders intercepting in the middle of sender and receiver and try to capture all the information transmitting from both parties. Diffie-Hellman key exchange technique [1] suffers from this type of problems. Intruders can act as sender and receiver simultaneously and try to steal secret session key at the time of exchanging key via public channel.

### 2.2. Methodology in MLPKG

This well known problem of middle man attack has been addressed in MLPKG where secret session key is not exchanged over public insecure channel. At end of neural weight synchronization strategy of both parties' generates identical weight vectors and activated hidden layer outputs for both the parties become identical. This identical output of hidden layer for both parties can be use as one time secret session key for secured data exchange.

## 3. Multilayer Perceptron based Key Generation System

A multilayer perceptron synaptic simulated weight based undisclosed key generation is carried out between recipient and sender. Figure1 shows multilayer perceptron based synaptic simulation system. Sender and receivers multilayer perceptron select same single hidden layer among multiple hidden layers for a particular session. For that session all other hidden layers goes in deactivated mode means hidden (processing) units of other layers do nothing with the incoming input. Either synchronized identical weight vector of sender and receivers' input layer, activated hidden layer and output layer becomes session key or session key can be form using identical output of hidden units of activated hidden layer. The key generation technique and analysis of the technique using random number of nodes (neurons) and the corresponding algorithm is discussed in the subsections 3.1 to 3.5 in details.

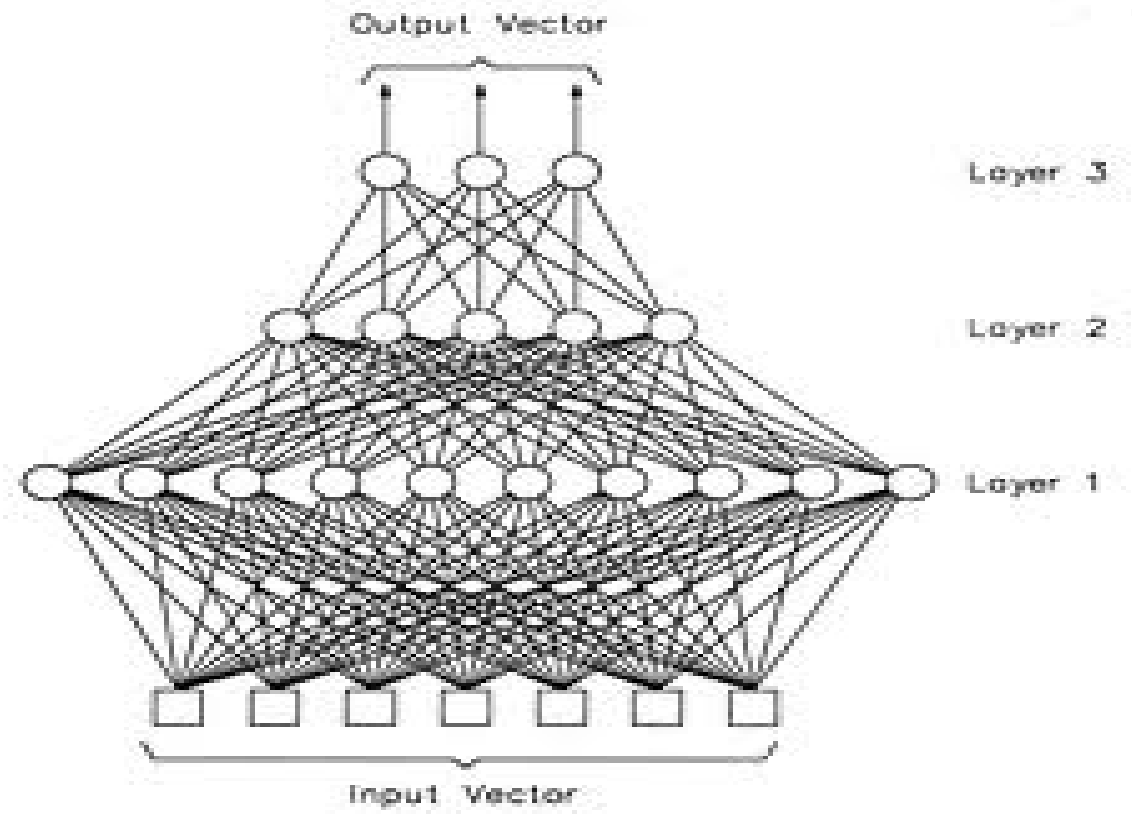

Figure 1. A Multilayer Perceptron with 3 Hidden Layers

Sender and receiver multilayer perceptron in each session acts as a single layer network with dynamically chosen one activated hidden layer and K no. of hidden neurons, N no. of input neurons having binary input vector, $x_{ij} \in \{-1,+1\}$, discrete weights, are generated from input to output, are lies between -L and +L, $w_{ij} \in \{-L, -L+1, \ldots, +L\}$. Where i = 1,…,K denotes the i$^{th}$ hidden unit of the perceptron and j = 1,…,N the elements of the vector and one output neuron. Output of the hidden units is calculated by the weighted sum over the current input values . So, the state of the each hidden neurons is expressed using (eq.1)

$$h_i = \frac{1}{\sqrt{N}} w_i x_i = \frac{1}{\sqrt{N}} \sum_{j=1}^{N} w_{i,j} x_{i,j} \qquad (1)$$

Output of the i$^{th}$ hidden unit is defined as

$$\sigma_i = \text{sgn}(h_i) \qquad (2)$$

But in case of $h_i$ = 0 then $\sigma_i$ = -1 to produce a binary output. Hence a, $\sigma_i$ = +1, if the weighted sum over its inputs is positive, or else it is inactive, $\sigma_i$ = -1. The total output of a perceptron is the product of the hidden units expressed in (eq. 2)

$$\tau = \prod_{i=1}^{K} \sigma_i \qquad (3)$$

### 3.1 Multilayer Perceptron Simulation Algorithm

**Input: -** Random weights, input vectors for both multilayer perceptrons.
**Output: -** Secret key through synchronization of input and output neurons as vectors.
**Method:-**

**Step 1.** *Initialization of random weight values of synaptic links between input layer and randomly selected activated hidden layer.*

*Where,* $w_{ij} \in \{-L, -L+1, ..., +L\}$ (4)

**Step 2.** *Repeat step 3 to 6 until the full synchronization is achieved, using Hebbian-learning rules.*

$$w^{+}_{i,j} = g\left(w_{i,j} + x_{i,j}\tau\Theta(\sigma_i\tau)\Theta\left(\tau^A\tau^B\right)\right) \quad (5)$$

**Step 3.** *Generate random input vector X. Inputs are generated by a third party or one of the communicating parties.*

**Step 4.** *Compute the values of the activated hidden neurons of activated hidden layer using (eq. 6)*

$$h_i = \frac{1}{\sqrt{N}} w_i x_i = \frac{1}{\sqrt{N}} \sum_{j=1}^{N} w_{i,j} x_{i,j} \quad (6)$$

**Step 5.** *Compute the value of the output neuron using*

$$\tau = \prod_{i=1}^{K} \sigma_i \quad (7)$$

*Compare the output values of both multilayer perceptron by exchanging the system outputs.*
*if Output (A) ≠ Output (B), Go to step 3*
*else if Output (A) = Output (B) then one of the suitable learning rule is applied only the hidden units are trained which have an output bit identical to the common output.*

Update the weights only if the final output values of the perceptron are equivalent. When synchronization is finally achieved, the synaptic weights are identical for both the system.

## 3.2 Multilayer Perceptron Learning rule

At the beginning of the synchronization process multilayer perceptron of A and B start with uncorrelated weight vectors $w_i^{A/B}$. For each time step K, public input vectors are generated randomly and the corresponding output bits $\tau^{A/B}$ are calculated. Afterwards A and B communicate their output bits to each other. If they disagree, $\tau^A \neq \tau^B$, the weights are not changed. Otherwise learning rules suitable for synchronization is applied. In the case of the Hebbian learning rule [10] both neural networks learn from each other.

$$w^{+}_{i,j} = g\left(w_{i,j} + x_{i,j}\tau\Theta(\sigma_i\tau)\Theta\left(\tau^A\tau^B\right)\right) \quad (8)$$

The learning rules used for synchronizing multilayer perceptron share a common structure. That is why they can be described by a single (eq. 4)

$$w^{+}_{i,j} = g\left(w_{i,j} + f\left(\sigma_i, \tau^A, \tau^B\right)x_{i,j}\right) \quad (9)$$

with a function $f\left(\sigma_i, \tau^A, \tau^B\right)$, which can take the values -1, 0, or +1. In the case of bidirectional interaction it is given by

$$f\left(\sigma_i, \tau^A, \tau^B\right) = \Theta\left(\sigma\tau^A\right)\Theta\left(\tau^A\tau^B\right)\begin{cases} \sigma & \textit{Hebbian learning} \\ -\sigma & \textit{anti-Hebbian learning} \\ 1 & \textit{Random walk learning} \end{cases} \tag{10}$$

The common part $\Theta\left(\sigma\tau^A\right)\Theta\left(\tau^A\tau^B\right)$ of $f\left(\sigma_i, \tau^A, \tau^B\right)$ controls, when the weight vector of a hidden unit is adjusted. Because it is responsible for the occurrence of attractive and repulsive steps [6].

### 3.3 Weight Distribution of Multilayer Perceptron

In case of the Hebbian rule (eq. 8), A's and B's multilayer perceptron learn their own output. Therefore the direction in which the weight $w_{i,j}$ moves is determined by the product $\sigma_i x_{i,j}$ . As the output $\sigma_i$ is a function of all input values, $x_{i,j}$ and $\sigma_i$ are correlated random variables. Thus the probabilities to observe $\sigma_i x_{i,j}$ = +1 or $\sigma_i x_{i,j}$ = -1 are not equal, but depend on the value of the corresponding weight $w_{i,j}$ [11, 13, 14, 15, 16].

$$P\left(\sigma_i x_{i,j} = 1\right) = \frac{1}{2}\left[1 + erf\left(\frac{w_{i,j}}{\sqrt{NQ_i - w_{i,j}^2}}\right)\right] \tag{11}$$

According to this equation, $\sigma_i x_{i,j} = \operatorname{sgn}(w_{i,j})$ occurs more often than the opposite, $\sigma_i x_{i,j} = -\operatorname{sgn}(w_{i,j})$ . Consequently, the Hebbian learning rule (eq. 8) pushes the weights towards the boundaries at -L and +L. In order to quantify this effect the stationary probability distribution of the weights for $t \to \infty$ is calculated for the transition probabilities. This leads to [11].

$$P\left(w_{i,j} = w\right) = P_0 \prod_{m=1}^{|w|} \frac{1 + erf\left(\frac{m-1}{\sqrt{NQ_i - (m-1)^2}}\right)}{1 - erf\left(\frac{m}{\sqrt{NQ_i - m^2}}\right)} \tag{12}$$

Here the normalization constant $\rho_0$ is given by

$$P_0 = \left(\sum_{w=-L}^{L} \prod_{m=1}^{|w|} \frac{1 + erf\left(\frac{m-1}{\sqrt{NQ_i - (m-1)^2}}\right)}{1 - erf\left(\frac{m}{\sqrt{NQ_i - m^2}}\right)}\right)^{-1} \tag{13}$$

In the limit $N \to \infty$ the argument of the error functions vanishes, so that the weights stay uniformly distributed. In this case the initial length of the weight vectors is not changed by the process of synchronization.

$$\sqrt{Q_i(t=0)} = \sqrt{\frac{L(L+1)}{3}} \tag{14}$$

But, for finite N, the probability distribution itself depends on the order parameter $Q_i$ Therefore its expectation value is given by the solution of the following equation:

$$Q_i = \sum_{w=-L}^{L} w^2 P(w_{i,j} = w) \tag{15}$$

### 3.4 Order Parameters

In order to describe the correlations between two multilayer perceptron caused by the synchronization process, one can look at the probability distribution of the weight values in each hidden unit. It is given by (2L + 1) variables.

$$P^i_{a,b} = P\left(w^A_{i,j} = a \wedge w^B_{i,j} = b\right) \tag{16}$$

which are defined as the probability to find a weight with $w^A_{i,j} = a$ in A's multilayer perceptron and $w^B_{i,j} = b$ in B's multilayer perceptron. In both cases, simulation and iterative calculation, the standard order parameters, which are also used for the analysis of online learning, can be calculated as functions of $P^i_{a,b}$ [12].

$$Q^A_i = \frac{1}{N} w^A_i w^A_i = \sum_{a=-L}^{L} \sum_{b=-L}^{L} a^2 P^i_{a,b} \tag{17}$$

$$Q^B_i = \frac{1}{N} w^B_i w^B_i = \sum_{a=-L}^{L} \sum_{b=-L}^{L} b^2 P^i_{a,b} \tag{18}$$

$$R^{AB}_i = \frac{1}{N} w^A_i w^B_i = \sum_{a=-L}^{L} \sum_{b=-L}^{L} abP^i_{a,b} \tag{19}$$

Then the level of synchronization is given by the normalized overlap between two corresponding hidden units

$$\rho^{AB}_i = \frac{w^A_i w^B_i}{\sqrt{w^A_i w^A_i}\sqrt{w^B_i w^B_i}} = \frac{R^{AB}_i}{\sqrt{Q^A_i Q^B_i}} \tag{20}$$

### 3.5 Hidden Layer as a Secret Session Key

At end of full weight synchronization process, weight vectors between input layer and activated hidden layer of both multilayer perceptron systems become identical. Activated hidden layer's output of source multilayer perceptron is used to construct the secret session key. This session key is not get transmitted over public channel because receiver multilayer perceptron has same identical activated hidden layer's output. Compute the values of the each hidden unit by

$$\sigma_i = \operatorname{sgn}\left(\sum_{j=1}^{N} w_{ij} x_{ij}\right) \quad \operatorname{sgn}(x) = \begin{cases} -1 & \text{if } x < 0, \\ 0 & \text{if } x = 0, \\ 1 & \text{if } x > 0. \end{cases} \tag{21}$$

For example consider 8 hidden units of activated hidden layer having absolute value (1, 0, 0, 1, 0, 1, 0, 1) becomes an 8 bit block. This 10010101 become a secret session key for a particular session and cascaded xored with recursive replacement encrypted text. Now final session key based encrypted text is transmitted to the receiver end. Receiver has the identical session key i.e. the output of the hidden units of activated hidden layer of receiver. This session key used to get the recursive replacement encrypted text from the final cipher text. In the next session both the machines started tuning again to produce another session key.

Identical weight vector derived from synaptic link between input and activated hidden layer of both multilayer perceptron can also becomes secret session key for a particular session after full weight synchronization is achieved.

## 4. CHARACTER CODE TABLE GENERATION

For plain text “tree” figure 2 shows corresponding tree representation of probability of occurrence of each character in the plain text. Character‘t’ and ‘r’ occur once and character ‘e’ occurs twice. Each character code can be generated by travelling the tree using preorder traversal. Character values are extracted from the decimal representation of character code. Left branch is coded as ‘0’ and that of right branch ‘1’. Table 1 shows the code and value of a particular character in the plain text. From the original tree mutated tree is derived using mutation. Figure 3, 4 and 5 are the mutated trees. After mutation new code values as obtained are tabulated in table 2. Tree having (n-1) intermediate nodes can generate $2^{n-1}$ mutated trees. In order to obtain unique value, the code length is added to the character if the value is identical in the table.

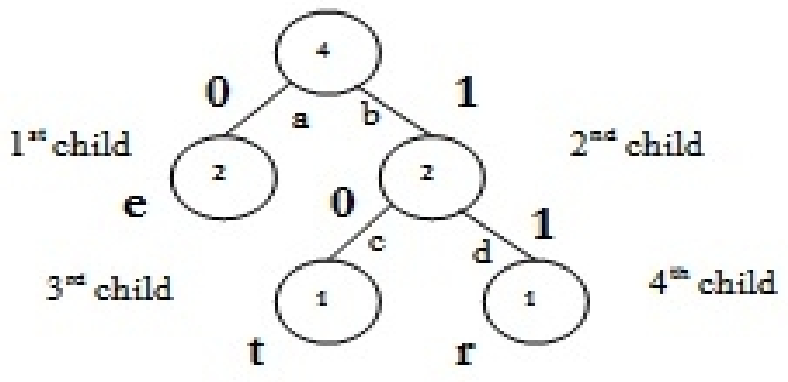

Figure 2. Character Code Tree

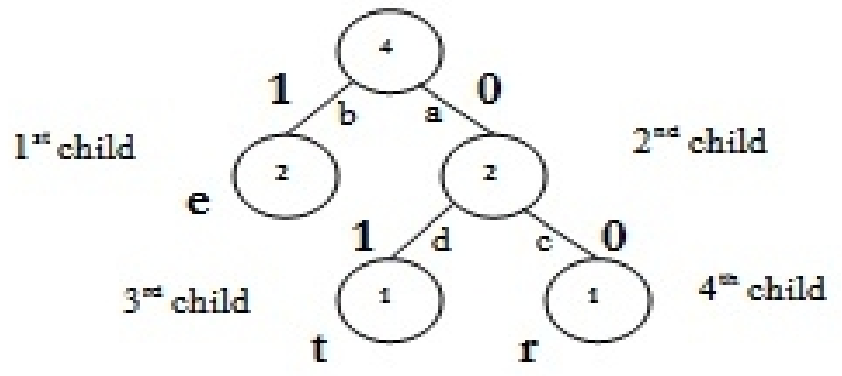

Figure 3. Swap the edges between a and b and between c and d.

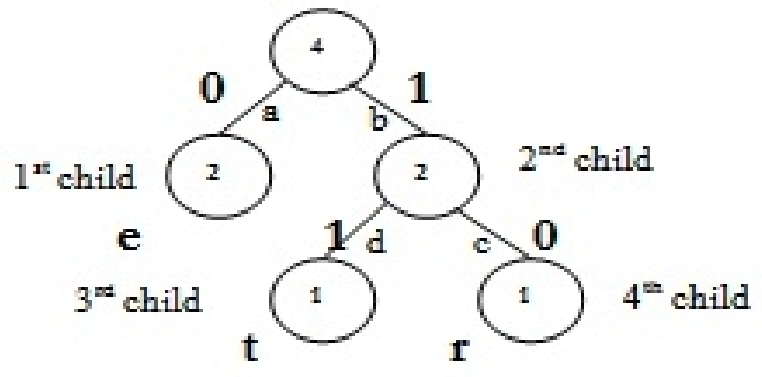

Figure 4. Swap the edges between c and d. Edges a and b get unaltered.

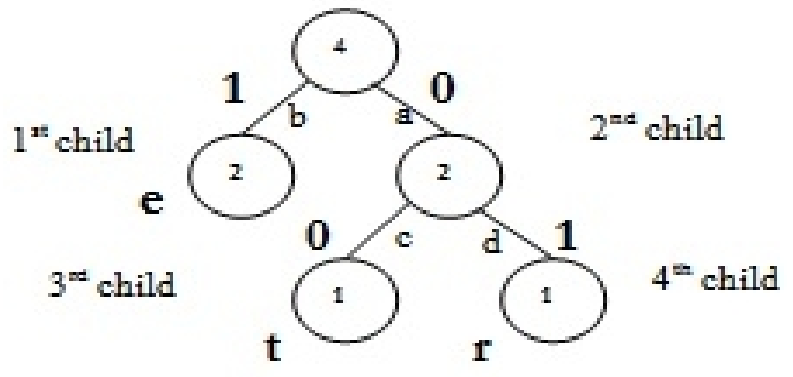

Figure 5. Swap the edges between a and b. Edges c and d get unaltered.

Table1. Code table

| Plain text | Code | Value |
|---|---|---|
| t | 10 | 2 |
| r | 11 | 3 |
| e | 0 | 0 |

Table2. Mutated code table

| Character | Code | Value | Code | Value | Code | Value |
|---|---|---|---|---|---|---|
| t | 01 | 1 | 11 | 3 | 00 | 0 |
| r | 00 | 0 | 10 | 2 | 01 | 1 |
| e | 1 | 2 | 0 | 0 | 1 | 2 |

# 5. RECURSIVE REPLACEMENT ENCRYPTION

**Step 1:** Decompose the source stream, say, into a finite number of blocks, each preferably of the same size, say, L.

**Step 2:** Calculate the total number of primes and nonprimes in the range of 0 to ($2^L$-1). Accordingly, find minimum how many bits are required to represent each of these two numbers.

Step 3 to be applied for all the blocks.

**Step 3:** For the block under consideration, calculate the decimal number corresponding to that. Say, it is D. Find out if D is prime or nonprime. If D is prime, the code value for that block is 1 and if not so, it is 0.In the series of primes or nonprimes (whichever be applicable for D) in the range of 0 to ($2^L$-1), find the position of D. Represent this position in terms of binary values. This is the rank of this block.

After repeating this step 3 for all the blocks, following steps are to be followed.

**Step 4:** Say, there are N number of blocks. In the target stream of bits, put all the N code values one by one starting from the MSB position. So, in the target stream, the first N bits are code values for N blocks.

**Step 5:** For putting all the rank values in the target stream, we are to start from the $N^{th}$ bit from the MSB position and then to come back bit-by-bit. Immediately after the $N^{th}$ bit, put the rank value of the $N^{th}$ block, followed by the rank value of the $(N-1)^{th}$ block, and so on. In this way, the rank value of the first block will be placed at the last.

**Step 6:** Combining all the code values as well as the rank values, if the total number of bits in the target stream is not a multiple of 8, then to make it so, at most 7 bits may have to be inserted. Insertion of these extra bits is to be started from the $(N+1)^{th}$ position. So, a maximum of 7 right shifting operations may have to be performed in the $(N+1)^{th}$ position, where that many 0's are inserted.

Now, MLP (multilayer perceptron) based secret session key is use to xor the recursive replacement encrypted stream. This MLP secret session key is use to xored with the same length first intermediate cipher text block to produce the first final cipher block (MLP secret session key XOR with same length cipher text). This newly generated block again xored with the immediate next block and so on. This chaining of cascaded xoring mechanism is performed until all the blocks get exhausted. If the last block size of intermediate cipher text is less than the require xoring block size (i.e. weight vector size) then this block is kept untouched.

# 6. EXAMPLE

Consider a stream S=1010100101010010 of only 16 bits. Apply these steps to obtain the target stream T corresponding to S using the recursive replacement technique.

Decompose S into four 4-bit blocks taking bits four by four from the MSB, which are $D_1$=1010, $D_2$=1001,$D_3$=0101 and $D_4$=0010. So, as per step 1, L=4.

Obtain the total number of primes in the range of 0 to $2^4$-1=15 is 6 (2, 3, 5, 7, 11, 13) and that of nonprimes is 10 (0, 1, 4, 6, 8, 9, 10, 12, 14, 15).

To represent the position of a prime number the number of bits required is 3, because since there are 6 primes, their positions range from 0 to 5, Similarly, to represent the position of a nonprime number the number of bits required is 4 because their positions range from 0 to 9 as there are 10 nonprime numbers.

Apply the next step 3 for blocks $D_1$, $D_2$, $D_3$ and $D_4$.

The decimal equivalent of $D_1$=1010 is 10, which is the $6^{th}$ nonprime. So, the code value of $D_1$ is $C_1$=0 and the rank is $R_1$=0110.

The decimal equivalent of $D_2$=1001 is 9, which is the $5^{th}$ nonprime. So, the code value of $D_2$ is $C_2$=0 and the rank is $R_2$=0101.

The decimal equivalent of $D_3$=0101 is 5, which is the $2^{nd}$ prime. So, the code value of $D_3$ is $C_3$=1 and the rank is $R_3$=010.

The decimal equivalent of $D_4$=0010 is 2, which is the $0^{th}$ prime. So, the code value of $D_4$ is $C_4$=1 and the rank is $R_4$=000.

To form the target stream, first we put all the code values one by one starting from the MSB position to get 0/0/1/1 and they are followed by the rank values of all the blocks starting from the last, i.e., 000/010/0101/0110. Here "/" works just as the separator.
Combining these code values and rank values we obtain 001100001001010110, a stream of length 18.

To make the length a multiple of 8, a block "000000" is to be inserted between the code values and the rank values, so that the stream 0011/000000/00001001010110 is formed.

Therefore corresponding to the 16-bit source stream S=1010100101010010, the 24-bit target stream is T=001100000000001001010110 as follows.

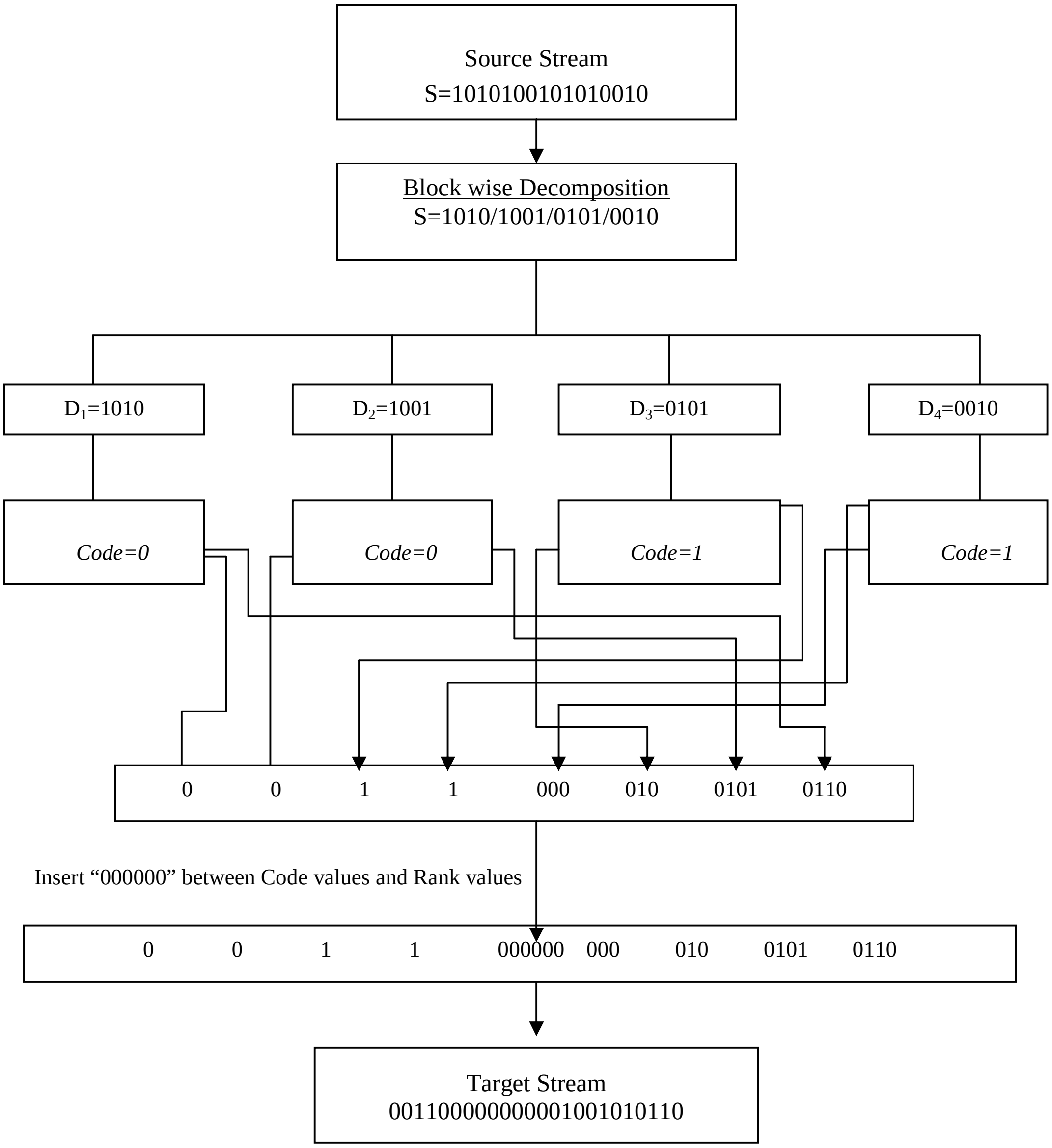

Figure 6. Pictorial Representation of Encrypting S=1010100101010010

In this way, we obtain the target stream as T=0100101001100101.Now further MLP generated key and recursively replacement encrypted text is use to finally encrypt the block.

## 7. RECURSIVE REPLACEMENT DECRYPTION

During decryption, it is to be noted that the receiver will take MLP secret session key. Then cascaded xoring operation is performed using MLP secret session key with the cipher text. The technique of performing xoring is same that was in encryption process. Then MLP secret session key is use to deciphering the outcomes of the previous step. Finally from the outcomes intermediate encrypted block (E) is extracted and now key is use to decipher the E to get the source stream.

Following are the set of steps to be followed for the purpose of recursive replacement decryption:

**Step 1:** Get the unique block length from the key. Say, it is L.

**Step 2:** Calculate the total number of blocks generated from the source stream of bits. The following does this calculation: Total Number of Blocks (B) = Source Stream Size / Unique Block Length, "/" denoting the integer division. So, the first B number of bits, starting from position 0 (MSB position) to position (B-1) in the encrypted stream denotes the code values of B blocks.

r

**Step 3:** Calculate the total number of primes in the range of 0 to ($2^L$-1). Say, it is P. Hence calculate how many maximum bits are required to express P in binary form. Say, it is X. Then X = $\lfloor \log_2 P \rfloor + 1$, where $\lfloor \log_2 P \rfloor$ denotes the integral part of $\log_2$ P.

**Step 4:** Calculate the total number of nonprimes in the range of 0 to ($2^L$-1). Say, it is Q. hence calculates how many maximum bits are required to express Q in binary form. Say, it is Y. Then Y = $\lfloor \log_2 Q \rfloor + 1$, where $\lfloor \log_2 Q \rfloor$ denotes the integral part of $\log_2$ Q. It is mentionable here that Q = $2^L$ – P.

**Step 5:** Consider the MSB. It is the code value of the first source block. If MSB=1, Consider the last block of X bits, convert the binary number represented by this block of bits into the corresponding decimal, Say, it is M. Mark this block as being processed. Find the $M^{th}$ prime number in the series of natural numbers (with the assumption that the position of the first prime number is 0, not 1). The L-bit binary number corresponding to the decimal prime number obtained in 2 is the first source block. Mark the MSB as being processed. If MSB=0,Consider the last block of Y bits; convert the binary number represented by this block of bits into the corresponding decimal, Say, it is M. Mark this block as being processed. Find the $M^{th}$ prime number in the series of natural numbers (with the assumption that the position of the first prime number is 0, not 1). The L-bit binary number corresponding to the decimal nonprime number obtained in 2 is the first source block. Mark the MSB as being processed.

**Step 6:** Repeat step 7 and step 8 for (B-1) number of times for the values of I ranging from 1 to (B-1) as there are (B-1) more blocks left to be considered. Set I = 1.

**Step 7:** Consider the $I^{th}$ bit from the MSB position. Let it be denoted by $T_I$. If $T_I$ = 1, Consider the first unprocessed block of P bits in the LSB-to-MSB direction, convert the binary number represented by this block of bits into the corresponding decimal, Say, it is M. Mark this block being processed. Find the $M^{th}$ prime number in the series of natural numbers (with the assumption that the position of the first prime number is 0, not 1). The L-bit binary number corresponding to the decimal prime number obtained in 2 is the $I^{th}$ source block. If $T_I$ = 0, Consider the first unprocessed block of Q bits in the LSB-to-MSB direction, convert the binary number represented by this block of bits into the corresponding decimal, Say, it is M. Mark this block being processed. Find the $M^{th}$ nonprime number in the series of natural numbers (with the assumption that the position of the first prime number is 0, not 1). The L-bit binary number corresponding to the decimal nonprime number obtained in 2 is the $I^{th}$ source block.

**Step 8:** Let I = I + 1.

**Step 9:** Concatenate all the blocks obtained so far in the sequence of their generation and this is the source stream. The length of the source stream is (L * B) and accordingly $L_T$ – (L * B) number of 0's in the positions between the code values and the target values in the target stream will remain being unmarked, as these 0's were inserted at the end of the encryption process; $L_T$ being considered as the length of the target stream.

## 8. EXAMPLE

We continue with the same example, where the target stream we obtained was T = 00110000000001001010110.
Now, from step 1, from the key we get the unique block length L = 4.

Following step 2, we obtain the total number of blocks as B = 16 / 4 = 4, as it is assumed to be known to the receiver that the source stream before being encrypted was of length 16 bits. Therefore in the encrypted stream, the first four bits are the code values of four blocks.

Following step 3, we calculate the total number of primes in the range of 0 to 15 (i.e., $2^4 – 1$), which is P = 6, and to represent it by a binary number the maximum number of bits needed is X = 3.

Similarly, following step 4, we calculate the total number of nonprimes in the range of 0 to 15 (i.e., $2^4 – 1$), which is P = 10, and to represent it by a binary number the maximum number of bits needed is Y = 4.

Now, following step 5, we find the MSB as 0, so that we are to consider the block of the last Y = 4 number of bits, which is 0110, the decimal of which is M = 6. So, we are to find the $6^{th}$ nonprime number in the series of natural numbers. It is 10 (assuming that 0 is the $0^{th}$ nonprime, 1 is the $1^{st}$ nonprime, and so on), the 4-bit binary of which is 1010. Hence the first source block is 1010.

Using step 6, we can say that step 7 and step 8 are to be repeated for 3 times as there are still 3 blocks left. Step 7 only does the job of moving from one block to another and, in fact, step 8 works in the same way as step 5. So, proceeding in the same way, we obtain the remaining blocks as 1001, 0101 and 0010.

Following step 9, we concatenate all the blocks in the same sequence of their generation to obtain the source stream S = 1010100101010010.

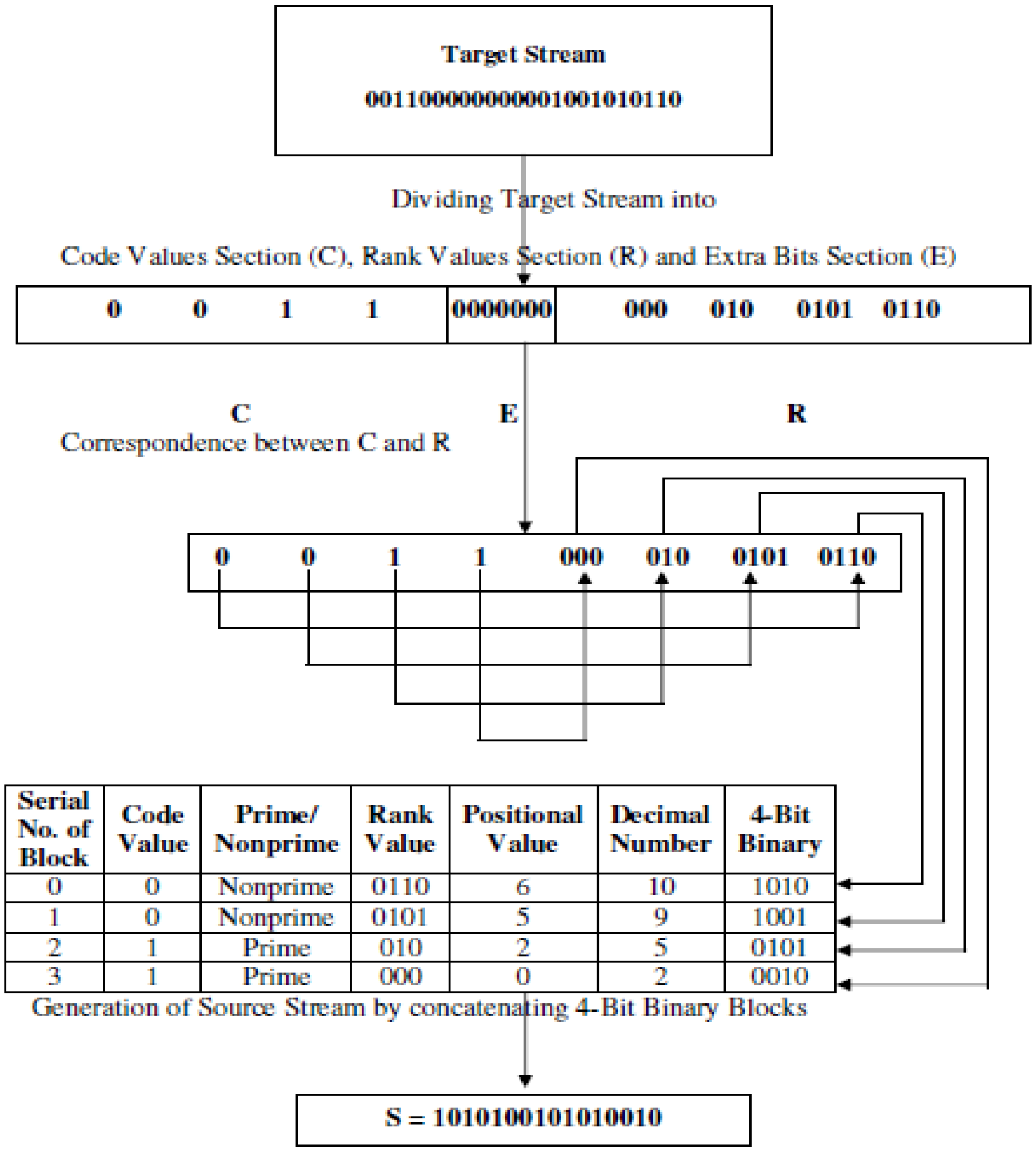

Figure 7. Pictorial Representation of Decrypting T = 001100000000001001010110

## 9. COMPLEXITY ANALYSIS

The complexity of the technique will be O(L), which can be computed using following three steps.

**Step 1.** To generate a MLP guided key of length N needs O(N) Computational steps. The average synchronization time is almost independent of the size N of the networks, at least up to N=1000.Asymptotically one expects an increase like O (log N).

**Step 2.** Complexity of the encryption technique is O(L).

***Step 2. 1.*** Recursive replacement of bits using prime nonprime recognition encryption process takes O(L).

***Step 2. 2.*** MLP based encryption technique takes O(L) amount of time.

**Step 3.** Complexity of the decryption technique is O(L).

***Step 3. 1.*** In MLP based decryption technique, complexity to convert final cipher text into recursive replacement cipher text T takes O(L).

***Step 3. 2.*** Transformation of recursive replacement cipher text T into the corresponding stream of bits S = $s_0$ $s_1$ $s_2$ $s_3$ $s_4 \ldots s_{L-1}$, which is the source block takes O(L) as this step also takes constant amount of time for merging $s_0$ $s_1$ $s_2$ $s_3$ $s_4 \ldots s_{L-1}$.

So, overall time complexity of the entire technique is O(L).

# 10. EXPERIMENT RESULTS

In this section the results of implementation of the proposed MLPKG technique has been presented in terms of encryption decryption time, Chi-Square test, source file size vs. encryption time along with source file size vs. encrypted file size. The results are also compared with existing RSA [1] technique, existing ANNRBLC [8] and NNSKECC [9].

.

Table 3. Encryption / decryption time vs. File size

| **Encryption Time (s)** | | | **Decryption Time (s)** | | |
|---|---|---|---|---|---|
| **Source Size (bytes)** | **MLPKG** | **NNSKECC [9]** | **Encrypted Size (bytes)** | **MLPKG** | **NNSKECC [9]** |
| 18432 | 6. 42 | 7.85 | 18432 | 6.99 | 7.81 |
| 23044 | 9. 23 | 10.32 | 23040 | 9.27 | 9.92 |
| 35425 | 14. 62 | 15.21 | 35425 | 14. 47 | 14.93 |
| 36242 | 14. 72 | 15.34 | 36242 | 15. 19 | 15.24 |
| 59398 | 25. 11 | 25.49 | 59398 | 24. 34 | 24.95 |

Table 3 shows encryption and decryption time with respect to the source and encrypted size respectively. It is also observed the alternation of the size on encryption.

In figure 8 stream size is represented along X axis and encryption / decryption time is represented along Y-axis. This graph is not linear, because of different time requirement for finding appropriate MLP key. It is observed that the decryption time is almost linear, because there is no MLP key generation process during decryption.

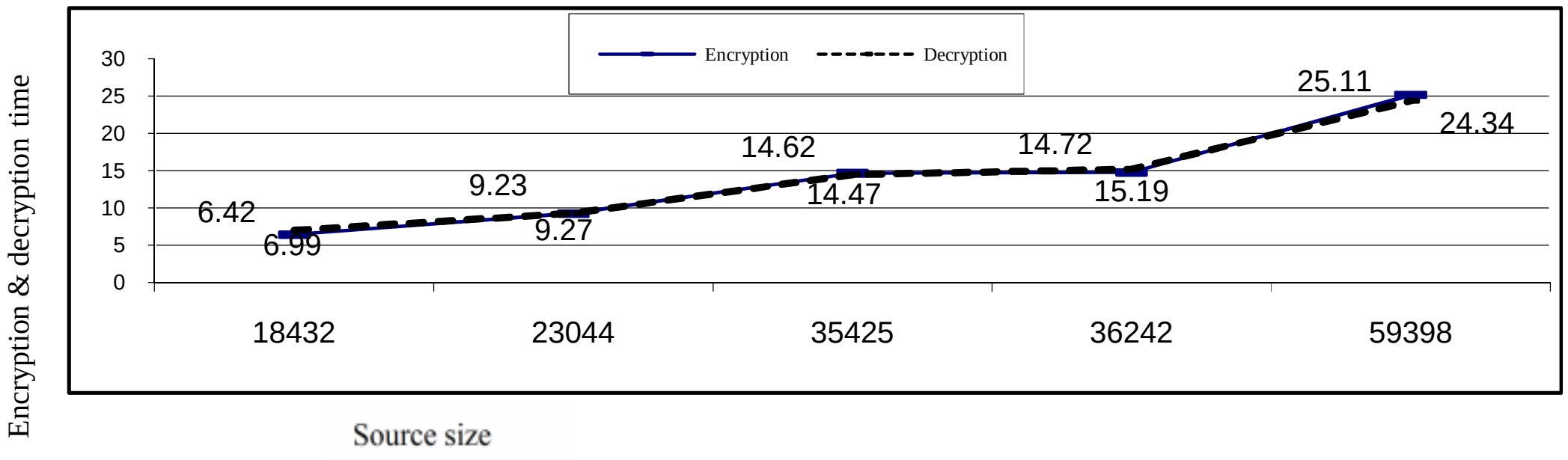

Figure 8. Source size vs. encryption time & decryption time

Table 4 shows Chi-Square value for different source stream size after applying different encryption algorithms. It is seen that the Chi-Square value of MLPKG is better compared to the algorithm ANNRBLC [8] and comparable to the Chi-Square value of the RSA algorithm.

Table 4. Source size vs. Chi-Square value

| **Stream Size (bytes)** | **Chi-Square value (TDES) [1]** | **Chi-Square value in (MLPKG)** | **Chi-Square value (ANNRBLC) [8]** | **Chi-Square value (RSA) [1]** |
|---|---|---|---|---|
| 1500 | 1228.5803 | 2856.2673 | 2471.0724 | 5623.14 |
| 2500 | 2948.2285 | 6582.7259 | 5645.3462 | 22638.99 |
| 3000 | 3679.0432 | 7125.2364 | 6757.8211 | 12800.355 |
| 3250 | 4228.2119 | 7091.1931 | 6994.6198 | 15097.77 |
| 3500 | 4242.9165 | 12731.7231 | 10572.4673 | 15284.728 |

Figure 9 shows graphical representation of table 4.

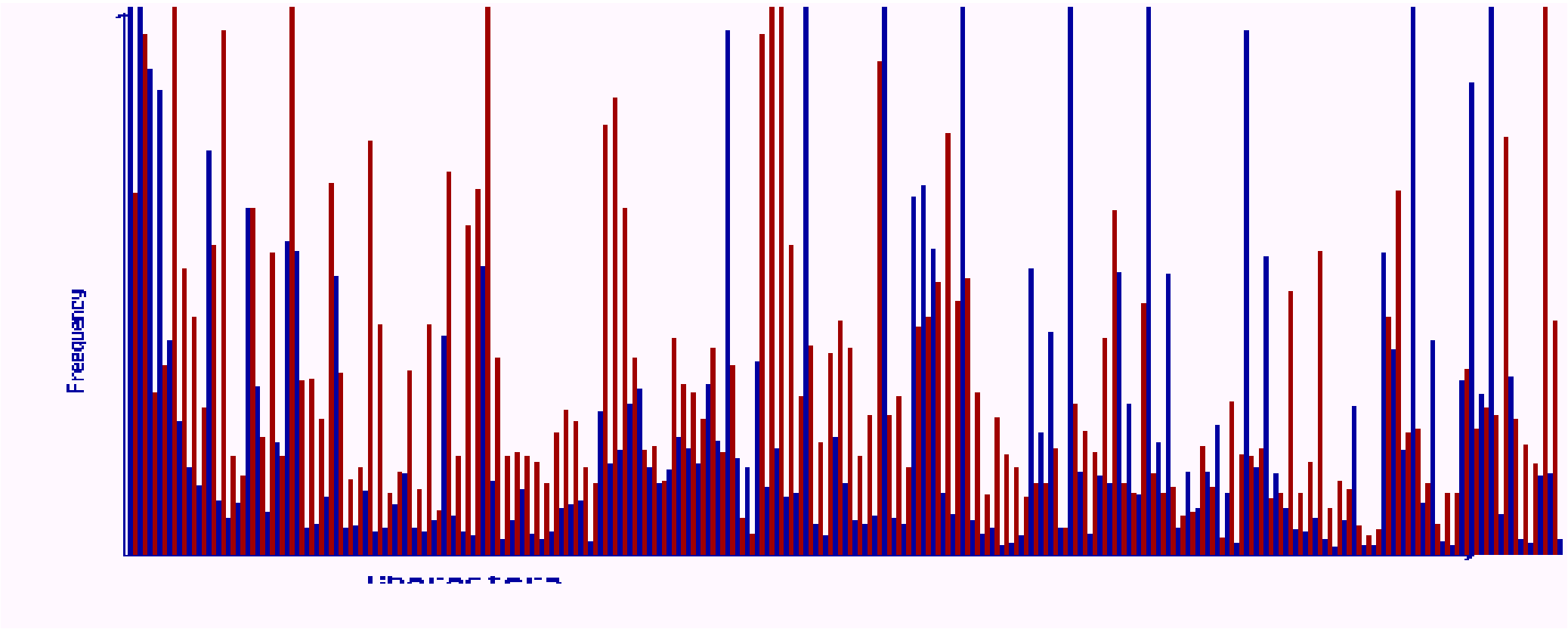

Figure 9. Source size vs. Chi-Square value

Table 5 shows total number of iteration  needed and number of data being transferred for MLP key generation process with different numbers of input(N) and activated hidden(H) neurons and varying synaptic depth(L).

Table 5. Data Exchanged and No. of Iterations For Different  Parameters Value

| No. of Input Neurons(N) | No. of Activated Hidden Neurons(K) | Synaptic Weight (L) | Total No. of Iterations | Data Exchanged (Kb) |
|---|---|---|---|---|
| 5 | 15 | 3 | 624 | 48 |
| 30 | 4 | 4 | 848 | 102 |
| 25 | 5 | 3 | 241 | 30 |
| 20 | 10 | 3 | 1390 | 276 |
| 8 | 15 | 4 | 2390 | 289 |

Following figure 10. Shows the snapshot of MLP key simulation process.

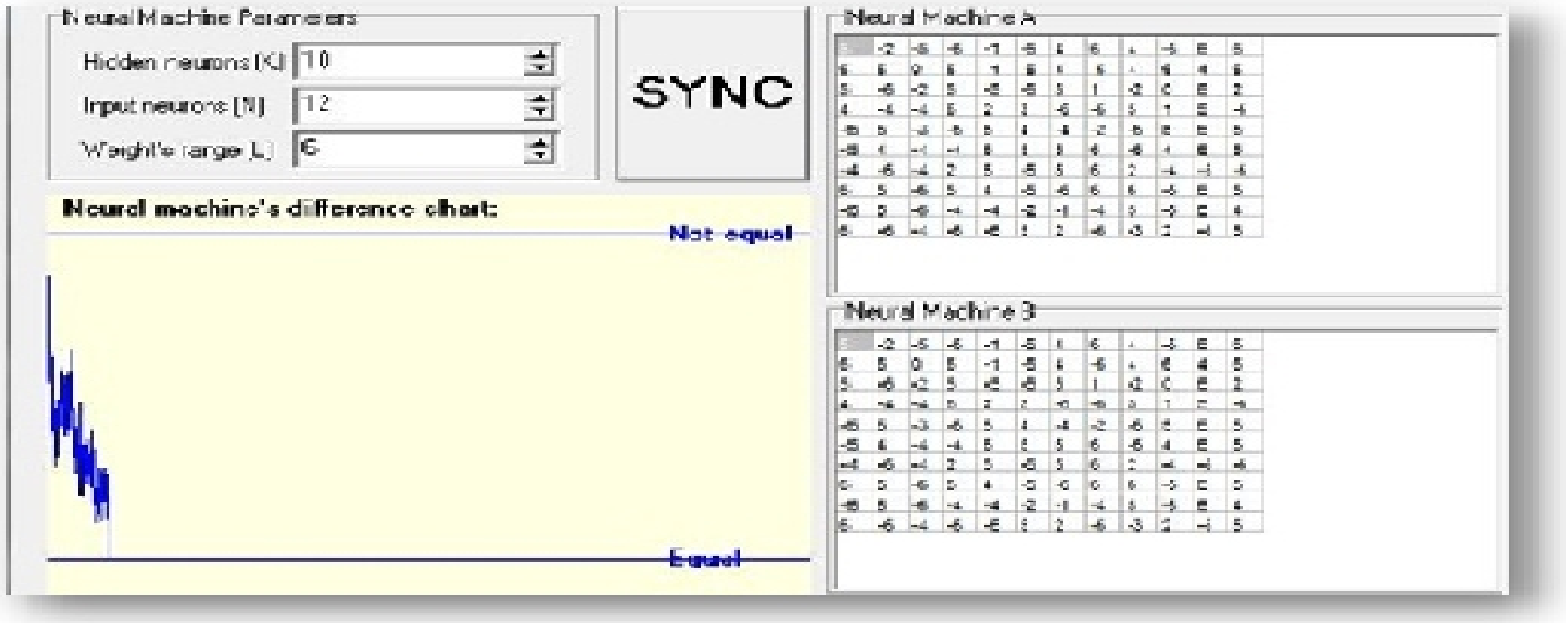

Figure 10. MLP Key Simulation Snapshot with N=12, K=10 and L=6

## 11. ANALYSIS OF RESULTS

From results obtained it is clear that the technique will achieve optimal performances. Encryption time and decryption time varies almost linearly with respect to the block size. For the algorithm presented, Chi-Square value is very high compared to some existing algorithms. A user input key has to transmit over the public channel all the way to the receiver for performing the decryption procedure. So there is a likelihood of attack at the time of key exchange. To defeat this insecure secret key generation technique a neural network based secret key generation technique has been devised. The security issue of existing algorithm can be improved by using MLP secret session key generation technique. In this case, the two partners A and B do not have to share a common secret but use their indistinguishable weights or output of activated hidden layer as a secret key needed for encryption. The fundamental conception of MLP based key exchange protocol focuses mostly on two key attributes of MLP. Firstly, two nodes coupled over a public channel will synchronize even though each individual network exhibits disorganized behaviour. Secondly, an outside network, even if identical to the two communicating networks, will find it exceptionally difficult to synchronize with those parties, those parties are communicating over a public network. An attacker E who knows all the particulars of the algorithm and records through this channel finds it thorny to synchronize with the parties, and hence to calculate the common secret key. Synchronization by mutual learning (A and B) is much quicker than learning by listening (E) [10]. For usual cryptographic systems, we can improve the safety of the protocol by increasing of the key length. In the case of MLP, we improved it by increasing the synaptic depth L of the neural networks. For a brute force attack using K hidden neurons, K*N input neurons and boundary of weights L, gives (2L+1)KN possibilities. For example, the configuration K = 3, L = 3 and N = 100 gives us 3*10253 key possibilities, making the attack unfeasible with today's computer power. E could start from all of the (2L+1)3N initial weight vectors and calculate the ones which are consistent with the input/output sequence. It has been shown, that all of these initial states move towards the same final weight vector, the key is unique. This is not true for simple perceptron the most unbeaten cryptanalysis has two supplementary ingredients first; a group of attacker is used. Second, E makes extra training steps when A and B are quiet [10]-[12]. So increasing synaptic depth L of the MLP we can make our MLP safe.

## 12. SECURITY ISSUE

The main difference between the partners and the attacker in MLP is that A and B are able to influence each other by communicating their output bits $\tau^A$ & $\tau^B$ while E can only listen to these messages. Of course, A and B use their advantage to select suitable input vectors for adjusting the weights which finally leads to different synchronization times for partners and attackers. However, there are more effects, which show that the two-way communication between A and B makes attacking the MLP protocol more difficult than simple learning of examples. These confirm that the security of MLP key generation is based on the bidirectional interaction of the partners. Each partener uses a seperate, but identical pseudo random number generator. As these devices are initialized with a secret seed state shared by A and B. They produce exactly the same sequence of input bits. Whereas attacker does not know this secret seed state. By increasing synaptic depth average synchronize time will be increased by polynomial time. But success probability of attacker will be drop exponentially Synchonization by mutual learning is much faster than learning by adopting to example generated by other network. Unidirectional learning and bidirectional synchronization. As E can't influence A and B at the time they stop transmit due to synchrnization. Only one weight get changed where, $\sigma_i$ = T. So, difficult to find weight for attacker to know the actual weight without knowing internal representation it has to guess.

# 13. FUTURE SCOPE & CONCLUSION

This paper presented a novel approach for generation of secret key proposed algorithm using MLP simulation. This technique enhances the security features of the key exchange algorithm by increasing of the synaptic depth L of the MLP. Here two partners A and B do not have to exchange a common secret key over a public channel but use their indistinguishable weights or outputs of the activated hidden layer as a secret key needed for encryption or decryption. So likelihood of attack proposed technique is much lesser than the simple key exchange algorithm.

Future scope of this technique is that this MLP model can be used in wireless communication. Some evolutionary algorithm can be incorporated with this MLP model to get well distributed weight vector.

## ACKNOWLEDGEMENTS

The author deep sense of gratitude to the DST, Govt. of India, for financial assistance through INSPIRE Fellowship leading for a PhD work under which this work has been carried out.

## REFERENCES

[1] Atul Kahate, Cryptography and Network Security, 2003, Tata McGraw-Hill publishing Company Limited, Eighth reprint 2006.

[2] Sarkar Arindam, Mandal J. K, "Artificial Neural Network Guided Secured Communication Techniques: A Practical Approach" LAP Lambert Academic Publishing ( 2012-06-04 ), ISBN: 978-3-659-11991-0, 2012

[3] Sarkar Arindam, Karforma S, Mandal J. K, "Object Oriented Modeling of IDEA using GA based Efficient Key Generation for E-Governance Security (OOMIG) ", International Journal of Distributed and Parallel Systems (IJDPS) Vol.3, No.2, March 2012, DOI : 10.5121/ijdps.2012.3215, ISSN : 0976 - 9757 [Online] ; 2229 - 3957 [Print]. Indexed by: EBSCO, DOAJ, NASA, Google Scholar, INSPEC and WorldCat, 2011.

[4] Mandal J. K., Sarkar Arindam, "Neural Session Key based Traingularized Encryption for Online Wireless Communication (NSKTE)", 2nd National Conference on Computing and Systems, (NaCCS 2012), March 15-16, 2012, Department of Computer Science, The University of Burdwan, Golapbag North, Burdwan –713104, West Bengal, India. ISBN 978- 93-808131-8-9, 2012.

[5] Mandal J. K., Sarkar Arindam, "Neural Weight Session Key based Encryption for Online Wireless Communication (NWSKE)", Research and Higher Education in Computer Science and Information Technology, (RHECSIT- 2012) ,February 21-22, 2012, Department of Computer Science, Sammilani Mahavidyalaya, Kolkata , West Bengal, India. ISBN 978-81- 923820-0-5,2012

[6] Mandal J. K., Sarkar Arindam, "An Adaptive Genetic Key Based Neural Encryption For Online Wireless Communication (AGKNE)", International Conference on Recent Trends In Information Systems (RETIS 2011) BY IEEE, 21-23 December 2011, Jadavpur University, Kolkata, India. ISBN 978-1-4577-0791-9, 2011

[7] Mandal J. K., Sarkar Arindam, "An Adaptive Neural Network Guided Secret Key Based Encryption Through Recursive Positional Modulo-2 Substitution For Online Wireless Communication (ANNRPMS)", International Conference on Recent Trends In Information Technology (ICRTIT 2011) BY IEEE, 3-5 June 2011, Madras Institute of Technology, Anna University, Chennai, Tamil Nadu, India. 978-1-4577-0590-8/11, 2011

[8] Mandal J. K., Sarkar Arindam, "An Adaptive Neural Network Guided Random Block Length Based Cryptosystem (ANNRBLC)", 2nd International Conference on Wireless Communications, Vehicular Technology, Information Theory And Aerospace & Electronic System Technology" (Wireless Vitae 2011) By IEEE Societies, February 28- March 03, 2011,Chennai, Tamil Nadu, India. ISBN 978-87-92329-61-5, 2011

[9] Mandal J. K., Sarkar Arindam, “Neural Network Guided Secret Key based Encryption through Cascading Chaining of Recursive Positional Substitution of Prime Non-Prime (NNSKECC)”, International Confference on Computing and Systems, ICCS – 2010, 19–20 November, 2010,Department of Computer Science, The University of Burdwan, Golapbag North, Burdwan – 713104, West Bengal, India.ISBN 93-80813-01-5, 2010
[10] R. Mislovaty, Y. Perchenok, I. Kanter, and W. Kinzel. Secure key-exchange protocol with an absence of injective functions. Phys. Rev. E, 66:066102,2002.
[11] A. Ruttor, W. Kinzel, R. Naeh, and I. Kanter. Genetic attack on neural cryptography. Phys. Rev. E, 73(3):036121, 2006.
[12] A. Engel and C. Van den Broeck. Statistical Mechanics of Learning. Cambridge University Press, Cambridge, 2001.
[13] T. Godhavari, N. R. Alainelu and R. Soundararajan “Cryptography Using Neural Network ” IEEE Indicon 2005 Conference, Chennai, India, 11-13 Dec. 2005.gg
[14] Wolfgang Kinzel and ldo Kanter, "Interacting neural networks and cryptography", Advances in Solid State Physics, Ed. by B. Kramer (Springer, Berlin. 2002), Vol. 42, p. 383 arXiv- cond-mat/0203011, 2002
[15] Wolfgang Kinzel and ldo Kanter, "Neural cryptography" proceedings of the 9th international conference on Neural Information processing(ICONIP 02).h
[16] Dong Hu "A new service based computing security model with neural cryptography"IEEE07/2009.J

**Arindam Sarkar**

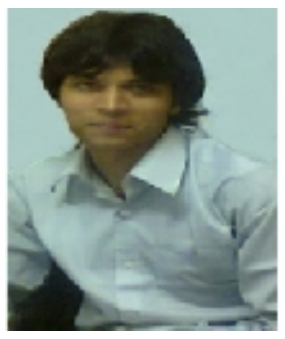

INSPIRE Fellow (DST, Govt. of India), MCA (VISVA BHARATI, Santiniketan, University First Class First Rank Holder), M.Tech (CSE, K.U, University First Class First Rank Holder). Total number of publications 8.

**Jyotsna Kumar Mandal**

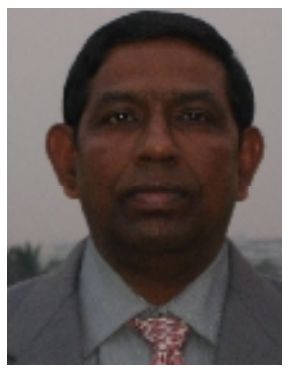

M. Tech.(Computer Science, University of Calcutta), Ph.D.(Engg., Jadavpur University) in the field of Data Compression and Error Correction Techniques, Professor in Computer Science and Engineering, University of Kalyani, India. Life Member of Computer Society of India since 1992 and life member of cryptology Research Society of India. Dean Faculty of Engineering, Technology & Management, working in the field of Network Security, Steganography, Remote Sensing & GIS Application, Image Processing. 25 years of teaching and research experiences. Eight Scholars awarded Ph.D. one submitted and 8 are pursuing. Total number of publications 230.